\documentclass[twocolumn,multicolumn,aps,prl,showpacs]{revtex4}
\usepackage{graphicx}
\usepackage{dcolumn}
\usepackage{bm}
\usepackage{color}
\usepackage{latexsym}
\usepackage[T1]{fontenc}
\usepackage{amsmath}

\input{epsf}

\begin{document}

\preprint{APS/123-QED}

\date{\today}

\title{Charge doping-induced quasiparticle scattering in 
iron-pnictide superconductors as probed by collective vortex pinning}

\author{Cornelis J. van der Beek$^{1}$, Marcin Konczykowski$^{1}$, Shigeru 
Kasahara$^{2}$, Takahito Terashima$^{2}$, Ryuji Okazaki$^{3}$, 
Takasada Shibauchi$^{3}$, Yuji Matsuda$^{3}$ }

\affiliation{\mbox{$^{1}$Laboratoire des Solides Irradi\'{e}s, CNRS UMR 7642 \& CEA-DSM-IRAMIS, Ecole Polytechnique, 91128 Palaiseau, France} \\
\mbox{$^{2}$Research Center for Low Temperature and Materials Sciences, Kyoto University, Sakyo-ku, Kyoto 606-8501, Japan} \\
\mbox{$^{3}$Department of Physics, Kyoto University, Sakyo-ku, Kyoto 
606-8502, Japan}\\
}

\begin{abstract}
Charge doping of iron-pnictide superconductors leads to collective 
pinning of flux vortices, whereas isovalent doping does not. Moreover, 
flux pinning in the charge-doped compounds is consistently described 
by the mean-free path fluctuations introduced by the dopant atoms, 
allowing for the extraction of the elastic quasiparticle scattering rate. The absence of scattering by dopant 
atoms in isovalently doped  BaFe$_{2}$(As$_{1-x}$P$_{x}$)$_{2}$ is 
consistent with the observation of a linear temperature dependence of 
the low-temperature penetration depth in this material.
\end{abstract}

\maketitle

With the advent of the superconducting iron pnictides 
\cite{Kamihara2008,Takahashi2008,Chen2008,Ren2008,Kito2008,Rotter2008} and 
chalcogenides, there are currently two classes of high temperature 
superconducting materials, the other being the cuprates. In both 
classes, superconductivity appears upon partial substitution of one  
or more elements of a magnetic parent material. Further 
substitution has the critical temperature $T_{c}$ go through a 
maximum, and back to zero on the overdoped side of the 
temperature-composition phase diagram.
Pnictides are specific in that this phenomenology may be 
induced either by charge doping or by 
isovalent substitutions. Known examples of the latter are the partial 
replacement of As by P \cite{Jiang2009,Shishido2010}, or Fe by Ru \cite{Sharma2010} in the BaFe$_{2}$As$_{2}$ ``122'' 
type materials, while charge doping is achieved by replacing O by F in the RBaFeO ``1111'' type materials (R is a 
rare earth element) \cite{Kamihara2008,Takahashi2008,Chen2008,Ren2008,Kito2008}, 
and Ba by K, or Fe by a transition metal 
ion in the 122's \cite{Rotter2008}. Introduction of either type of substitution 
causes important changes in band structure 
\cite{Shishido2010,Liu2010,Brouet2010}; charge 
doping cannot be reduced to a rigid shift of the Fermi level in 
these multi-band superconductors. Finally, dopant atoms act as 
scattering impurities, which in the weak scattering (Born) limit 
would couple quasiparticle excitations on different Fermi surface 
sheets, with possible repercussions 
\cite{Onari2009,Mishra2009,Kontani2010,Glatz2010} for the type of superconducting 
order parameter that may be realized  \cite{Mazin2008,Kuroki2008,Kuroki2009}, as well as for the diminishing 
$T_{c}$ in the overdoped region of the phase diagram due to 
pair-breaking \cite{Kogan2009}.

In this Letter, we focus on the latter aspect of the problem, and 
argue that charged dopant atoms act as scattering impurities for 
quasi-particles, while isovalent substitutions do not. The approach 
used is that of pinning of the vortex lattice by the impurities. The 
dimension of the vortex cores, of the order of the coherence length 
$\xi \sim 2$ nm, implies a high sensitivity not only to extrinsic but also to intrinsic 
disorder in superconductors. Thus, in electron-doped 
PrFeAsO$_{1-y}$, NdFeAsO$_{1-x}$F$_{x}$, and 
Ba(Fe$_{1-x}$Co$_{x}$)$_{2}$As$_{2}$, as in hole-doped 
Ba$_{1-x}$K$_{x}$Fe$_{2}$As$_{2}$, the 
critical current density $j_{c}$ is consistently described in terms of collective 
pinning  mediated by spatial fluctuations of the quasi-particle mean free 
path \cite{Blatter94,Thuneberg84,vdBeek91}. The impurity density accounting 
for pinning closely corresponds to the dopant atom concentration. 
Analysis of $j_{c}$ 
allows one to estimate the scattering cross-section and scattering phase angle 
$\delta_{0}$ of the defects, which turns out to be best described by the Born limit. 
On the other hand, isovalently doped 
BaFe$_{2}$(As$_{1-x}$P$_{x}$)$_{2}$ is characterized by a monotonous  power--law 
decrease of $j_{c}$ as function of magnetic flux density $B$, 
indicative of pinning solely by nm-scale disorder \cite{vdBeek2010}.


Critical current densities of single crystalline PrFeAsO$_{0.9}$ (with 
 $T_{c} \sim 35 $ K) \cite{vdBeek2010,Ishikado2009,Hashimoto2009,Okazaki2009}, 
NdFeAsO$_{0.9}$F$_{0.1}$  ($T_{c} \sim 36$ K)
\cite{vdBeek2010,Prozorov2008ii,Pribulova2009,Kacmarcik2009},
Ba$_{0.45}$K$_{0.55}$Fe$_{2}$As$_{2}$ ($T_{c} \sim 34$ K) \cite{Hashimoto2009ii}, 
and BaFe$_{2}$(As$_{x}$P$_{1-x}$)$_{2}$  \cite{Kasahara2009,Hashimoto2010}
were obtained from local measurements of the magnetic flux density  perpendicular 
to the crystal surface, $B_{\perp}$, and the flux density gradient $dB_{\perp}/dx \propto j_c$. 
Previous work has shown $j_{c}$ of superconducting iron pnictide 
crystals to be spatially inhomogeneous \cite{vdBeek2010}. While a 
\em global \rm measurement of the  \em average \rm flux density over the 
crystal surface, or of the magnetic moment of the entire crystal, may 
result in a spurious temperature dependence $j_{c}(T)$, local measurements 
do not have this shortcoming. Local $j_{c}$ values in applied fields 
up to 50 mT were obtained from magneto-optical imaging of the flux 
density \cite{vdBeek2010,Dorosinskii92}. 
Measurements in fields up to 2 T were performed using micron-sized 
Hall probe arrays, tailored in a pseudomorphic GaAlAs/GaAs heterostructure \cite{Okazaki2009}. 
The 10 Hall sensors of the array, spaced by 20 $\mu$m,  had an active area 
of $3 \times 3$ $\mu$m$^{2}$, while an 11th sensor was used for the 
measurement of the applied field. 

\begin{figure}[tb]
\includegraphics[width=0.45\textwidth]{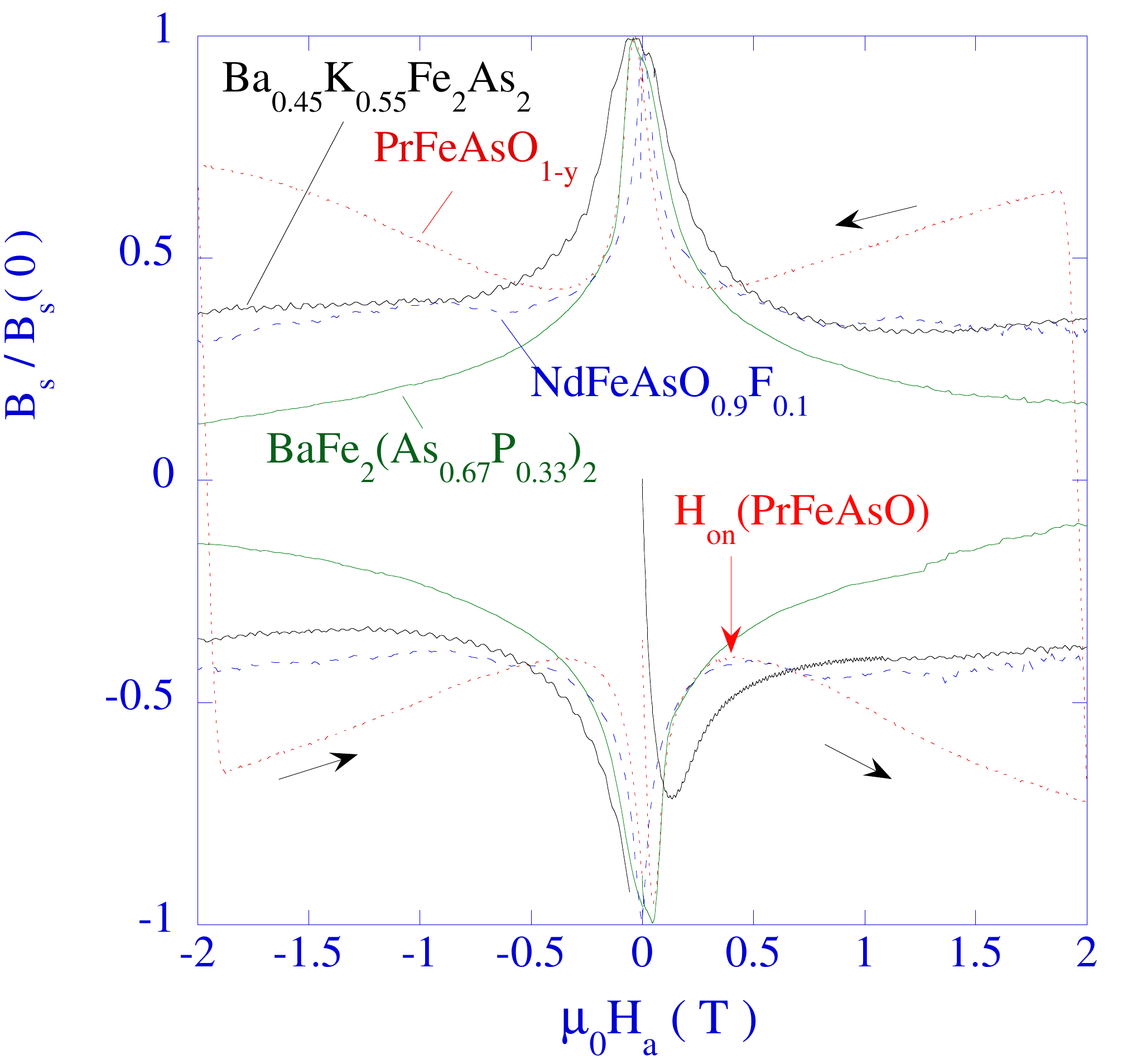}
\caption{(Color online) Normalized hysteresis loops of the local 
``self--field'', measured 
on the center of the top surfaces of PrFeAsO$_{1-y}$, NdFeAsO$_{0.9}$F$_{0.1}$,
Ba$_{0.45}$K$_{0.55}$Fe$_{2}$As$_{2}$, and 
BaFe$_{2}$(As$_{0.67}$P$_{0.33}$)$_{2}$  single crystals, at reduced 
temperature $T/T_{c} = 0.3$. Arrows 
indicate the direction in which the cycles are traversed. }
\label{fig:Loops}
 \vspace{-3mm}
\end{figure}

Figure~\ref{fig:Loops} shows hysteresis cycles of the local 
``self--field'' $B_{s} \equiv B_{\perp} - \mu_{0}H_{a}$  (where $\mu_{0} \equiv 4\pi 
\times 10^{-7}$ Hm$^{-1}$) versus 
$H_{a}$ for a variety of iron pnictide superconductors, at the same 
reduced temperature $
T/T_{c} = 0.2$. A salient feature of 
the hysteresis loops is the presence of a pronounced peak at small 
field. In the ``1111'' family of iron pnictide superconductors
\cite{vdBeek2010}, as in the Ba(Fe$_{1-x}$Co$_{x}$)$_{2}$As$_{2}$ 
\cite{Prozorov2008,Yamamoto2009,Prozorov2009} and 
Ba$_{1-x}$K$_{x}$Fe$_{2}$As$_{2}$ \cite{Yang2008} ``122'' 
superconductors, this peak is 
superposed on a field-independent contribution. At higher 
fields, the hysteresis loop width in these compounds increases
again, at a field $H_{on}$, the result of a structural change of the vortex 
ensemble \cite{vdBeek2010}. On the contrary, in isovalently doped 
BaFe$_{2}$(As$_{0.67}$P$_{0.33}$)$_{2}$, the hysteresis loop width shows 
a monotonous decrease.

\begin{figure}[tb]
\hspace{-2mm}
\includegraphics[width=0.65\textwidth]{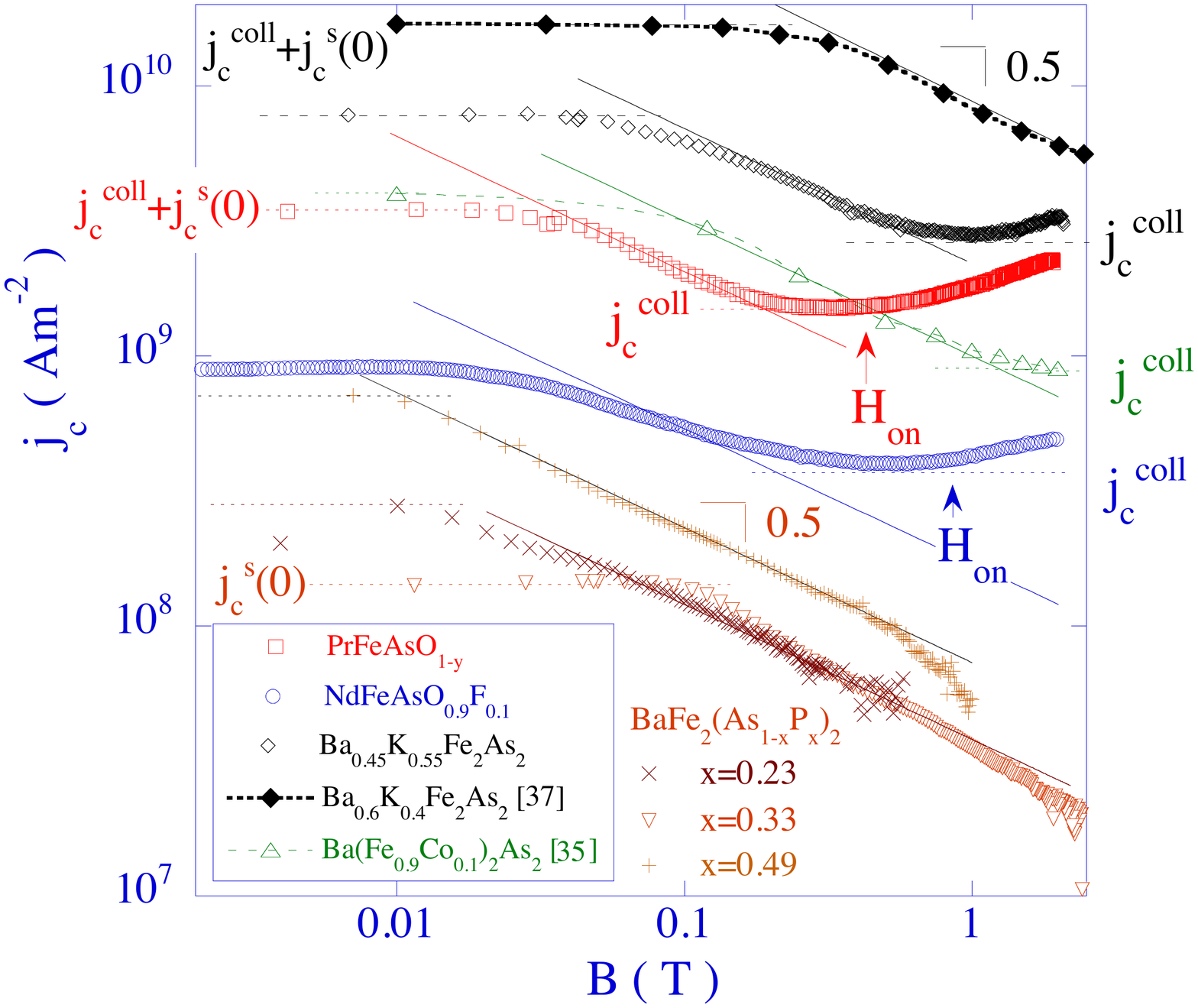}
\vspace{-29pt}
\caption{(Color online) Critical current density as function of magnetic flux 
density for PrFeAsO$_{1-y}$ (\color{red}$\Box$\color{black}), 
NdFeAsO$_{0.9}$F$_{0.1}$ (\color{blue}$\circ$\color{black}), Ba(Fe$_{0.9}$Co$_{0.1}$)$_{2}$As$_{2}$ 
(\color{green}$\triangle$\color{black}) \protect\cite{Yamamoto2009}, 
Ba$_{1-x}$K$_{x}$Fe$_{2}$As$_{2}$  (\protect\rotatebox{45}{\protect\rule{3.5pt}{3.5pt}}, \protect\raisebox{0.5pt}{$\diamond$}), 
and BaFe$_{2}$(As$_{1-x}$P$_{x}$)$_{2}$ single crystals of different $x$, at  $T/T_{c} = 0.3$. Drawn lines indicate the power-law dependence of pinning contribution from sparse pointlike defects. }
\label{fig:jc-B}
\end{figure}

\begin{figure}[h]
\includegraphics[width=0.68\textwidth]{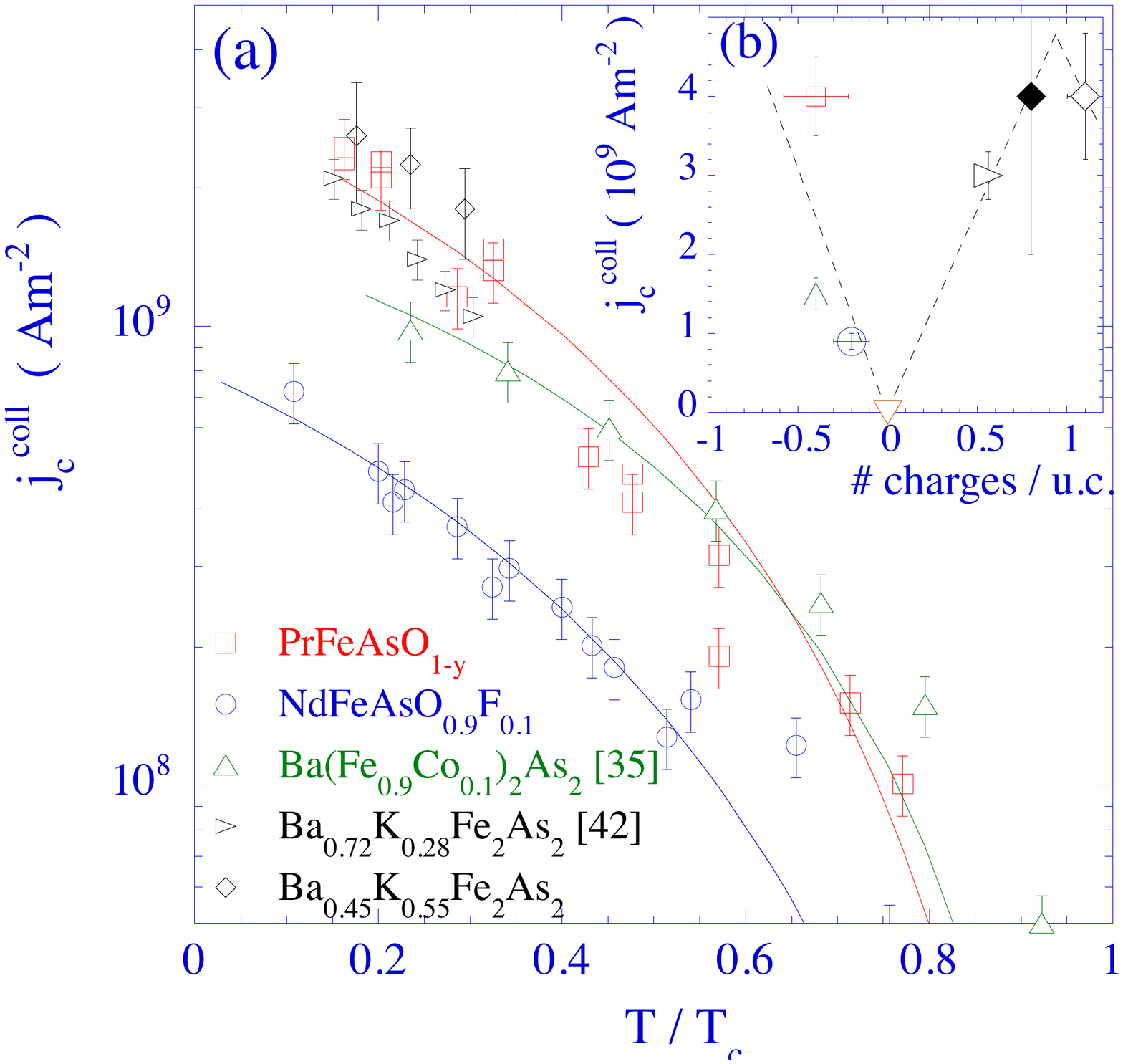}
\vspace{-32pt}
\caption{(Color online) (a) 
Collective pinning contribution 
$j_{c}^{coll}$ in the charge-doped compounds  PrFeAsO$_{1-y}$, 
NdFeAsO$_{0.9}$F$_{0.1}$, 
Ba$_{1-x}$K$_{x}$Fe$_{2}$As$_{2}$, 
Ba(Fe$_{0.9}$Co$_{0.1}$)$_{2}$As$_{2}$ 
\protect\cite{Yamamoto2009}, and  
BaFe$_{2}$(As$_{1-x}$P$_{x}$)$_{2}$ single crystals. 
Drawn lines are fits to 
Eq.~(\protect\ref{eq:1DCC-jc}). (b) The value of 
$j_{c}^{coll}$, extrapolated  to $T/T_c = 0.1$, as function of the number of dopant charges per 
unit cell. $j_{c}^{coll} = 0$ for BaFe$_{2}$(As$_{1-x}$P$_{x}$)$_{2}$ ($\bigtriangledown$). A point for Ba$_{0.6}$K$_{0.4}$Fe$_{2}$As$_{2}$ \cite{Yang2008} has been added (\protect\rotatebox{45}{\protect\rule{4pt}{4pt}},  not shown in Fig. \protect\ref{fig:jc-T}a).}
\label{fig:jc-T}
\vspace{-6mm}
\end{figure}

Figure~\ref{fig:jc-B} shows the field dependence of $j_c$ 
of five iron-pnictide compounds. For the four compounds measured in the 
present study, $j_{c} = 2 \mu_{0}^{-1} dB_{\perp}/dx$ is extracted using the Bean model 
 \cite{Brandt96}, while 
data for Ba(Fe$_{0.9}$Co$_{0.1}$)$_{2}$As$_{2}$ and 
Ba$_{0.6}$K$_{0.4}$Fe$_{2}$As$_{2}$ are taken from 
Refs.~\cite{Yamamoto2009} and \cite{Yang2008}, respectively. In what 
follows, we describe $j_{c}$ as the superposition of two 
contributions, $j_{c}^{coll}$ and $j_{c}^{s}(B)$. The former 
accounts for the constant hysteresis width at higher fields, and 
the latter for the low-field peak. In all materials, the 
peak has the shape of a plateau, 
\begin{equation}
    j_{c}(0) = j_{c}^{coll} + j_{c}^{s}(0), 
    \label{eq:together}
\end{equation}
followed by a power-law decrease, such that 
\begin{equation}
    j_{c}(B) = j_{c}^{coll} + j_{c}^{s}(B) \sim j_{c}^{coll} + A B_{\perp}^{-\beta}       
    \label{eq:together-B}
\end{equation}
with $0.5 < \beta < 0.63$. The behavior of $j_{c}^{s}(B)$ is
that expected for vortex pinning by sparse pointlike defects of radius 
larger than $\xi$, which are inevitably present 
in any real, imperfect crystal 
\cite{Ovchinnikov91,vdBeek2002,Blatter2004}.  An analysis of
data on the RFeAsO iron pnictides has shown that spatial variations of 
the average dopant atom density on a (large) scale of several dozen nm, 
leading to concomittant modulations of $T_{c}$, account for the measured 
magnitude and temperature dependence of $j_{c}^{s}$ \cite{vdBeek2010}.  
Oppositely, the field-independent $j_{c}^{coll}$ is 
attributed to atomic scale fluctuations of the dopant atom positions (collective 
pinning) \cite{Blatter94}. The different field dependence of the critical current 
contributions (Fig.~\ref{fig:jc-B}) allows one to extract both as 
function of temperature. Figure~\ref{fig:jc-T}a shows the 
$T$--dependence of  $j_{c}(0)=j_{c}^{coll} + j_{c}^{s}(0)$. The lower panel \ref{fig:jc-T}b shows 
the field-independent contribution $j_{c}^{coll}$, 
non-zero in the charge-doped compounds, 
but absent for all investigated isovalent 
substitutions $x$ in BaFe$_{2}$(As$_{1-x}$P$_{x}$)$_{2}$.

We quantitatively describe $j_{c}^{coll}$ by treating the dopant atoms as point defects 
responsible for quasi-particle scattering. The elementary pinning 
force of such defects can be written  
\begin{equation}
f_{p} \sim 0.3 g(\rho_{D}) \varepsilon_{0} \left( \sigma_{tr}/\pi 
\xi^{2}\right) \left( \xi_{0}/\xi \right),
\end{equation}
where $\sigma_{tr} = (2\pi/k_{F}^{2}) \sin^{2} \delta_{0} = \pi D_{v}^{2}$ is the 
transport scattering cross-section, $k_{F}$ is the Fermi wavevector, 
$D_{v}$ is the effective range of the potential, and $g(\rho_{D})$ is the Gor'kov 
function \cite{Thuneberg84,vdBeek91,Blatter94}. The disorder 
parameter $\rho_{D} = \hbar v_{F}/2\pi T_{c} l\sim \xi_{0}/l$, with 
$v_{F}$ the Fermi velocity, $l = (n_{d}\sigma_{tr})^{-1}$ the 
quasi-particle mean free path, $n_d$ the defect density, and $\xi_{0} \approx 1.35 \xi(0)$ the (temperature-independent) 
Bardeen-Cooper-Schrieffer coherence length \cite{Thuneberg84,vdBeek91,Blatter94}. 
The critical current arises from the local density fluctuations of 
the defects, and is therefore determined by the second moment of 
the elementary pinning force, $\langle f_{p}^{2} \rangle$. 
Applying the theory of collective pinning \cite{vdBeek91,Blatter94}, 
it reads \cite{vdBeek2010}
\begin{equation}
    j_{c}^{coll}  \approx  j_{0}   
    \left[ \frac{0.01 n_{d} \sigma_{tr}^{2}}{\varepsilon_{\lambda} \xi} \left( \frac{\xi_{0}}{\xi} \right)^{2} \right]^{2/3} 
    \propto   \left[\frac{\lambda(0)}{\lambda(T)}\right]^{2}  \left( 
    1 - \frac{T}{T_{c}} \right)^{\alpha},
    \label{eq:1DCC-jc}
\end{equation}
where $j_{0} \equiv  \Phi_{0} / \sqrt{3} \pi \mu_{0} \lambda_{ab}^{2} \xi$ 
is the depairing current density, and $\varepsilon_{\lambda} \equiv 
\lambda_{ab}/\lambda_{c}$ the penetration depth anisotropy. 
Eq.~(\ref{eq:1DCC-jc}) does not depend on the symmetry of the 
superconducting ground state. However, the exponent $\alpha \sim 2$ 
does depend on the different weight that distinct 
Fermi surface sheets have in contributing to superconductivity in 
different compounds. Here it is treated as a phenomenological 
parameter, 
obtained from the ratio of $ab$-plane and 
$c$-axis penetration depths in the different compounds 
\cite{Okazaki2009,Pribulova2009}, 
 while $\lambda(0)/\lambda(T)$ was published in 
Refs.~\cite{Hashimoto2009,Okazaki2009} and \cite{Pribulova2009}.

\begin{table}[b]
\vspace{1mm}
\begin{tabular}{lcccccccccc}
    \hline
    \hline
    Compound & $k_{F}$ (\AA$^{-1}$) &  $\xi_{0}$ ( nm ) & $n_{d}$ ( 
    nm$^{-3}$) & $\sigma_{tr}$ ( \AA$^{2}$ ) & $D_{v}$ (\AA) &  
    $n_{d}D_{v}^{3}$ & $n_{d}\xi_{0}^{3}$ &  $\sin \delta_{0}$ & $\Gamma$ ( meV ) & $l$ ( nm ) \\
    \hline
    PrFeAsO$_{1-y}$                         & 0.33 & 2.4 & 1.5 & 6.7  & 1.46  & $5\times  10^{-3}$    & 21 &  0.3(2) &  10 & 10 \\
 
    NdFeAsO$_{0.9}$F$_{0.1}$                & 0.33 & 3.3 & 1.5 & 2.5  &  0.9  & $1\times 10^{-3}$     & 54 &  0.2 & 4 & 25 \\
   
    Ba(Fe$_{0.9}$Co$_{0.1}$)$_{2}$As$_{2}$  & 0.25 & 1.6 &  2  & 2.5 &  0.9  & $1.5\times 10^{-3}$   & 8  &  0.17 & 5 & 20 \\
    
    Ba$_{0.72}$K$_{0.28}$Fe$_{2}$As$_{2}$ \protect\cite{Wang2010} & 0.4  & 2.4 &  2.8  &  1.5  &  0.7  & $1\times 10^{-3}$   & 38  &  0.1(4) & 3 &  23 \\
    
    Ba$_{0.6}$K$_{0.4}$Fe$_{2}$As$_{2}$  \protect\cite{Yang2008}   & 0.5  & 2.2 &  4  & $2.5 \pm  1.3$  &  $0.8 \pm 0.2$ & $2\times 10^{-3}$   & 43  &  0.2 & 8 &  10 \\
    
    Ba$_{0.45}$K$_{0.55}$Fe$_{2}$As$_{2}$     & 0.5  & 2.2 &  5.5  &    1.5  &  0.7  & $2\times 10^{-3}$   & 59  &  0.2 &  10 &  12 \\

    BaFe$_{2}$(As$_{0.67}$P$_{0.33}$)$_{2}$ & 0.3 \protect\cite{Shishido2010} & 1.6 & 3.3 &  $< 1.5\times 10^{-2}$ & $< 0.1$ & $< 1\times 10^{-6}$ & 14 &  -- &  -- & --  \\
    \hline
    \hline
\end{tabular}
\caption{\mbox{Fundamental parameters and contribution of dopant disorder to the elastic scattering 
parameters of various  } \\ \mbox{ iron pnictide 
superconductors, as deduced from the collective pinning part of 
the critical current density, $j_{c}^{coll}$.}}
\label{table:scattering}
\end{table}

Figure~\ref{fig:jc-T}a shows that the temperature dependence of 
$j_{c}^{coll}$ is very well described by Eq.~(\ref{eq:1DCC-jc}). In PrFeAsO$_{1-y}$, its magnitude is accurately reproduced by inserting 
$\sigma_{tr} = \pi D_{v}^{2}$, with the oxygen ion radius $D_{v} = 
1.46$ \AA  \hspace{1pt} 
and $n_{d} \approx 1.5 \times 10^{27}$ m$^{-3}$. This  corresponds 
to the oxygen vacancy concentration at the doping level, $y \sim 0.1$.  
Thus, the collective pinning contribution  to the critical current density of the PrFeAsO$_{1-y}$ compound is 
well described by the  quasi-particle mean-free path fluctuation 
mechanism of Refs.~\cite{Blatter94,Thuneberg84,vdBeek91}.  The same holds true for NdFeAsO$_{1-x}$F$_{x}$ and 
Ba(Fe$_{0.9}$Co$_{0.1}$)$_{2}$As$_{2}$. If one takes defect densities 
corresponding to the dopant atom concentration, $n_{d} \sim  1.5 \times 
10^{27}$, $\sim 1 \times 10^{27}$, and $4\times 10^{27}$ m$^{-3}$  respectively,
very satisfactory fits to $j_{c}^{coll}(T)$ can be obtained using the scattering 
cross-sections of Table~\ref{table:scattering}. As far as 
Ba$_{1-x}$K$_{x}$Fe$_{2}$As$_{2}$ is concerned,  $j_{c}^{s}$ exceeds 
$j_{c}^{coll}$ by more than an order of magnitude, which
prohibits a reliable determination of the latter at high temperature. 
Therefore, we only consider the low-$T$ magnitude of $j_{c}^{coll}$ 
for this compound. The dopant atom densities lead to   
values $n_{d}\xi_{0}^{3}$ that are largely in excess of unity, justifying 
 the collective pinning approach \cite{Blatter94}, and $n_{d}D_{v}^{3} \ll 1$, which means that 
background scattering is irrelevant -- each defect can be considered  
independent \cite{Thuneberg84}.

The correlation between the collective pinning contribution $j_{c}^{coll}$ to the critical current density and the nominal number of dopant charges per unit cell  
is shown in Fig.~\ref{fig:jc-T}b. Since there are two formula units per unit cell, the number of dopant charges is defined as twice the product of the dopant valency and the doping fraction $x$.
In BaFe$_{2}$(As$_{1-x}$P$_{x}$)$_{2}$, $j_{c}^{coll}$  is unmeasurably small,  implying a qualitative difference  between charge--doped and isovalently substituted  iron  pnictide superconductors. 

The collective pinning effect  implies that the core charge on the doping impurities is incompletely screened, consistent with a Thomas-Fermi screening length,  $\sim 1$ nm. From
Table~\ref{table:scattering}, one sees that  scattering by 
charged dopants in the iron-pnictide superconductors is rather in 
the  Born limit ($\sin \delta_{0} \ll 1$).  Therefore, if  so-called $s_\pm$ superconductivity \cite{Mazin2008,Kuroki2008,Kuroki2009}, with a sign change of the order parameter on different Fermi  \newpage \noindent surface sheets, is realized in these materials, the scattering would be detrimental \cite{Onari2009,Mishra2009,Glatz2010}.  A crude assessment of the pair-breaking effect can be made using the 
 quasi-particle scattering rates $\Gamma \sim n_{d}  [\pi 
N_{n}(0)]^{-1} \sin^{2} \delta_{0}$ (with $N_{n}(0) = m k_{F}/\pi^{2}\hbar^{2}$ the 
density of states and $m$ the electronic mass)  estimated from 
experiment (see Table~\ref{table:scattering}). These turn out to be of 
the order $\Gamma \sim 1.7 T_{c}$ for NdFeAsO$_{0.9}$F$_{0.1}$ and 
Ba$_{0.78}$K$_{0.22}$Fe$_{2}$As$_{2}$, and $\Gamma \sim 4 T_{c}$ for 
the other charge-doped compounds. When inserted in the Abrikosov-Gor'kov relation, 
$\ln(T_{c}/T_{c0}) = \Psi(\frac{1}{2}) - \Psi(\frac{1}{2} + \Gamma / 
2\pi k_{B} T_{c})$ (with $\Psi$ the digamma function) 
\cite{AG}, this implies that
the $T_{c}$'s of the charge-doped pnictides would be reduced by a factor 2-5 from a hypothetical $T_{c0}$ 
in the absence of disorder. Moreover, superconductivity should become gapless at impurity densities 
much less than the actual dopant concentration \cite{Glatz2010}. 
Within the hypothesis of nodal extended $s$-wave 
superconductivity \cite{Mishra2009}, the obtained scattering rates imply a 
$T_{c}/T_{c0} \sim 0.5-0.7$. Finally, for fully gapped, 
non-sign changing multiband $s$-wave superconductivity (''$s_{++}$''), 
the dopant atoms or vacancies are not pair-breaking, 
and their effect is the averaging of the gap components on 
different Fermi surface sheets.
A different situation occurs in 
 BaFe$_{2}$(As$_{1-x}$P$_{x}$)$_{2}$ 
material, which is characterized by  the  \em absence \rm of quasi-particle 
scattering. Isoelectronic dopant disorder is benign to superconductivity with order 
parameter nodes, as this was observed by penetration depth 
measurements \cite{Hashimoto2010}. Furthermore, our analysis shows no clear distinction 
between scattering centers in the FeAs planes (such as Co), and 
out-of-plane defects, which attests to the three-dimensional 
nature of superconductivity in the low-field limit due to the 
contribution of the more dispersive hole-like sheets \cite{Okazaki2009}, 
centered on the $\Gamma$-point \cite{Singh2008}.  

In conclusion, it is shown that the analysis of collective vortex pinning provides clues as to microscopic scattering mechanisms in superconductors. In that, the analysis of the critical current density adds another transport property to the spectrum of techniques available for the quantification of disorder effects in superconductors. Applied to iron pnictide superconductors, we find strong indications that charged atomic sized defects, including dopant atoms, are responsible for quasi-particle scattering in the Born limit. The presence of such defects in charge-doped pnictides should have consequences for suggested $s_{\pm}$ superconductivity in these  materials. On the other hand, isovalently doped 
BaFe$_{2}$(As$_{1-x}$P$_{x}$)$_{2}$, which has a superconducting 
ground state with gap nodes \cite{Hashimoto2010}, is characterized by 
the absence of such quasi-particle scattering.

We thank V. Mosser, H. Eisaki, and P.C. 
Canfield for providing the Hall sensor arrays, the 
PrFeAsO$_{1-y}$ crystals, and the NdFeAs(O,F) crystals, respectively. 
This work was supported by the French National Research agency, under 
grant ANR-07-Blan-0368 "Micromag", by KAKENHI from JSPS, and by
Grant-in-Aid for the Global COE program ``The Next Generation
of Physics, Spun from Universality and Emergence'' from MEXT, Japan. 
R.O. was supported by the JSPS Research Foundation for Young Scientists.


\end{document}